\documentclass[%
preprint,
 amsmath,amssymb,
 aps,
]{revtex4-1}

\usepackage{graphicx}
\usepackage{dcolumn}
\usepackage{bm}

\begin{document}


\title{Phase Matching Quantum Key Distribution based on Single-Photon Entanglement}

\author{Wei Li$^{1,2,3}$}
\author{Le Wang$^{1,2}$}
\author{Shengmei Zhao$^{1,2}$}%
 \email{zhaosm@njupt.edu.cn}
\affiliation{$^{1}$Nanjing University of Posts and Telecommunications, Institute of Signal Processing and Transmission, Nanjing, 210003, China.}%
\affiliation{$^{2}$Nanjing University of Posts and Telecommunications, Key Lab Broadband Wireless Communication and Sensor Network, Ministy of Education, Nanjing, 210003, China.}%
\affiliation{$^{3}$National Laboratory of Solid State Microstructures, Nanjing University, Nanjing 210093, China.}%


\date{\today}

\begin{abstract}

Two time-reversal quantum key distribution (QKD) schemes are the quantum entanglement based device-independent (DI)-QKD and measurement-device-independent (MDI)-QKD. The recently proposed twin field (TF)-QKD, also known as phase-matching (PM)-QKD, has improved the key rate bound from $O\left( \eta \right )$ to $O\left( \sqrt {\eta} \right )$ with $\eta$ the channel transmittance. In fact, TF-QKD is a kind of MDI-QKD but based on single-photon detection. In this paper, we propose a different PM-QKD based on single-photon entanglement, referred to as single-photon entanglement-based phase-matching (SEPM)-QKD, which can be viewed as a time-reversed version of the TF-QKD. Detection loopholes of the standard Bell test, which often occur in DI-QKD over long transmission distances, are not present in this protocol because the measurement settings and key information are the same quantity which is encoded in the local weak coherent state. We give a security proof of SEPM-QKD and demonstrate in theory that it is secure against all collective attacks and beam-splitting attacks. The simulation results show that the key rate enjoys a bound of $O\left( \sqrt {\eta} \right )$ with respect to the transmittance. SEPM-QKD not only helps us understand TF-QKD more deeply, but also hints at a feasible approach to eliminate detection loopholes in DI-QKD for long-distance communications.

\end{abstract}

\pacs{Valid PACS appear here}
\maketitle

\section{Introduction}

\par Quantum key distribution (QKD), a secure communication method to enabling a secret random number string to be shared by two well-separated parties, says Alice and Bob, has been proven to be robust against channel attacks and against the power of quantum computation\cite{ekert1991quantum,pirandola2019advances}. The random number string, known only to Alice and Bob, can be used to encrypt messages transmitted between them. In theoretical research, the work has focused on the security of QKD taking into consideration the imperfections of actual devices\cite{lo1999unconditional,shor2000simple,mayers2001unconditional,gottesman2004security,xu2019quantum}. In practical applications, research on the extractable key rate has been categorized as focusing on improving the key rate, such as decoy state protocols\cite{wang2005beating,lo2005decoy,ma2005practical,wang2013three,wang2015free}, asymmetric coding\cite{yin2016measurement,zhou2016making}, higher dimensional systems\cite{wang2018high,bouchard2018experimental,mower2013high,canas2017high,ding2017high,etcheverry2013quantum}, and parameter optimization\cite{ma2012statistical,xu2014protocol,yu2015statistical,cai2009finite}, or focusing on improving the key transmission distance\cite{lo2012measurement,liu2013experimental,yin2016measurement}.

\par In addition to the recent satellite QKD scheme\cite{liao2017satellite}, the current mainstream QKD is based on photon transmission over optical fiber. For a given QKD scheme, the factors that determine the key rate and transmission distance are the error rate and the transmittance $\eta$. In the initial stage of QKD research, a single-photon was used as the carrier of quantum information and secret key rate was bounded to $O\left(\eta\right )$\cite{pirandola2009direct,pirandola2017fundamental}, which is equal to the maximum probability of successful detection of a single-photon state. The measurement-device-independent (MDI)-QKD proposed latter is based on the correlation measurement of a two-photon state and closed all detection loopholes\cite{lo2012measurement,braunstein2012side}. Regardless of the technical challenges of practical experiments, the transmission distance of MDI-QKD almost doubled compared with BB84. However, the transmittance for a single-photon in MDI-QKD is unchanged, and so the key rate is still bounded by $O\left( \eta \right )$. 

\par In Lucamarini et al. (2018) the twin-field (TF)-QKD\cite{lucamarini2018overcoming}, also known as phase-matching (PM)-QKD by Ma et al.\cite{ma2018phase}, was proposed to improve the key rate and was shown to beat the PLOB bound\cite{pirandola2017fundamental}. TF-QKD and PM-QKD are essentially identical, the former reflects what states are used to carry the keys, and the latter reflects how the keys are generated. After that, some variants of TF-QKD have been proposed, such as the sending or not sending protocol by Wang et al.\cite{wang2018twin,yu2019sending} and removing of phase randomization and postselection in the coding mode by cui et al.\cite{cui2019twin}. TF-QKD is a single-photon version of MDI-QKD\cite{lin2018simple,yin2019measurement}, in which a single count is used to extract the quantum key. In TF-QKD, the information carrier is no longer a single photon but a weak coherent field or wave state with definite phase and amplitude\cite{yin2019measurement}. Independent coherent states with locked global phase can interfere with each other, so they can be used in phase matching to extract keys. A weak coherent state can be approximated as a coherent superposition of a vacuum state and a single photon state. The detection probability has a $\sqrt{\eta}$ dependence on the channel transmittance, which leads to a bound for key rate of $O\left( \sqrt{\eta} \right )$. Because $\eta$ is a quantity less than 1, this protocol further enhances the transmission distance of rate keys in optical fibers. 

\par Indeed, MDI-QKD itself may be regarded as a time-reversed version of an entanglement-based device-independent (DI)-QKD\cite{bennett1992quantum,rubenok2013real}, and therefore conclude that TF-QKD is a time-reversed version of the single-photon entanglement-based DI-QKD. Over 30 years ago, scientists proposed and experimentally verified the existence of single-photon entanglement and confirmed the Bell inequalities for quantum correlations in different forms\cite{tan1991nonlocality,banaszek1999testing,lee2000quantum,babichev2004homodyne,van2005single,morin2013witnessing}. Subsequently, single-photon-entanglement-based DI-QKD was proposed in which the key is extracted according to whether Alice or Bob has detected that photon\cite{kamaruddin2015device}.However, this work did not attract much attention, let alone the relationship between this protocol and TF-QKD. In our previous work\cite{li2019wave}, we proposed confirming Bell inequalities for single-photon entanglement from joint measurements in wave space-the conjugate space of the photon number space. As a new carrier of quantum information, the wave state has similar properties to the weak coherent state; both can be viewed as a coherent superposition of a vacuum state and a single-photon state. In this paper, we propose single-photon entanglement-based phase-matching (SEPM)-QKD, which is actually a TF-QKD with quantum entanglement. In this protocol, single-photon entanglement provides the quantum link in the communications between Alice and Bob, who choose the two groups of phases to encode the key. Monitoring Eve's eavesdropping is performed by detecting violations of Bell inequality. Security proofing against collective attacks and beam-splitting attacks is thereby established. We also compare the key rate of SEPM-QKD with the wave-state-based QKD, as for TF(PM)-QKD and single-photon-based QKD, like the BB84- and MDI-QKD protocols. 

\section{Theory of single photon entanglement}

\par The physical basis of SEPM-QKD is the detection of single-photon entanglement in wave space, (Fig.1). When a third-party Charlie directs a single-photon state onto an optical beam splitter, the photon states at the two output ports may be regarded as an entangled state of a vacuum state $\left | 0 \right \rangle$ and a single-photon state $\left | 1 \right \rangle$ in the two path modes\cite{tan1991nonlocality}
\begin{equation}
\left | \Psi_{A,B} \right \rangle= \frac{\sqrt{2}}{2}\left[ e^{i\theta}\left | 1 \right \rangle_{A} \left | 0 \right \rangle_{B}+\left | 0 \right \rangle_{A} \left | 1 \right \rangle_{B} \right],
\end{equation}
where $e^{i\theta}$ is the accumulated phase difference between the two arms. Because the production of a single-photon from a single-photon source is probabilistic, a heralded single-photon source can be used to increase the proportion of effective counting. Equation (1) is a representation of single-photon entanglement in photon-number space. Based on the wave-particle duality in quantum mechanics, it is convenient to call the conjugate space of this photon number space the wave space.  Applying a two-dimensional Fourier transformation, we obtain single-photon entanglement in the conjugate space
\begin{equation}
\left | \Psi_{A,B} \right \rangle= \frac{\sqrt{2}}{2} e^{i\left( \theta-\alpha \right)} [ \left | \alpha_{A} \right \rangle_{w} \left | \left(\alpha-\theta\right)_{B} \right \rangle_{w}
-\left | \left(\alpha+ \pi \right )_{A} \right \rangle_{w} \left | \left( \alpha-\theta+\pi \right )_{B} \right \rangle_{w}  ],
\end{equation}
where states with a subscript $w$ denote wave states, $\alpha$ and $\alpha-\theta$ each with a value ranging from 0 to $2\pi$ denotes the phase characterizing Alice's and Bob's wave state. The pair of orthogonal bases states in wave space are
\begin{equation}
\begin{split}
\left | \alpha \right \rangle_{w}=&\frac{\sqrt{2}}{2} \left [ \left | 0 \right \rangle+e^{i\alpha}\left | 1 \right \rangle \right ],\\
\left | \alpha+\pi \right \rangle_{w}=&\frac{\sqrt{2}}{2} \left [ \left | 0 \right \rangle-e^{i\alpha}\left | 1 \right \rangle \right ].
\end{split}
\end{equation}
It is these states that are used to distribute the quantum correlation between Alice and Bob. 

\par Next, we analyze single-photon entanglement in wave space. Here we refer to the photon states $\left | \alpha \right \rangle_{w}$, $\left | \alpha+\pi \right \rangle_{w}$  as the $Z$ basis if $\alpha=0$, and $\left | \alpha \right \rangle_{w}$, $\left | \alpha+\pi \right \rangle_{w}$ as the $Y$ basis if $\alpha=\frac{\pi}{2}$; then the states $ \left | 0 \right \rangle$ and $ \left | 1 \right \rangle$ belong to the $X$ basis. We can see that the entanglement between Alice and Bob in the wave space is entirely determined by the value of phase $\theta$. Because the value of $\alpha$ is any real number, then, if we set the value of $\theta$ to zero, the initial single-photon entangled state is the Bell state $\left | \Phi_{A,B}^{-} \right \rangle_{w}$, which is rotationally symmetric in the $ZY$ plane. It should be noted that $\theta$ can also be set to other values, but the way they generate keys will change accordingly.

\par In our previous work, we demonstrated that a wave state could be measured through interference with a reference weak coherent state\cite{li2019wave}, as shown in the measurement device at the sites of Alice and Bob. Assuming that the weak coherent states selected by Alice and Bob are $\left | \gamma e^{i\alpha} \right \rangle$ and $\left | \gamma e^{i\beta} \right \rangle$ where $\gamma$ is a small amplitude far less than 1, the weak coherent state has the approximate form
\begin{equation}
\left | \gamma e^{i\alpha} \right \rangle \approx  \left | 0 \right \rangle +\gamma e^{i \alpha} \left | 1 \right \rangle +O\left( \gamma \right) \left | \text{other} \right \rangle,
\end{equation}
where $\alpha$ and $\beta$ are the phase values of the wave states of Alice and Bob, the state $\left | \text{other} \right \rangle$ with an infinitesimal amplitude is a coherent combination of Fock states whose photon number is greater than or equal to 2. Taking into account the transmittance of a single photon $\eta$ in the optical channel, the dependence of the measurement results on measurement settings $\alpha$ and $\beta$ reads
\begin{equation}
p\left(A_{i},B_{j} \right)=\frac{\gamma^{2}\eta}{4}\left[1+\left ( -1 \right )^{i+j} \cos \left( \alpha-\beta\right ) \right]+\frac{\gamma^{4}}{4},
\end{equation}
where $i,j\in \left \{ 1,2 \right \}$ are the ordinal numbers the single-photon detectors of Alice and Bob. The first term represents the wave-like correlation between Alice and Bob, while the second term represents the particle-like correlation between them\cite{li2019wave}. If the intensity of the weak coherent field $\gamma^{2}$ is far less than the transmittance $\eta$, then the second term on the right-hand side of Eq. (4) may be omitted. Now the time reversal relationship between SEPM-QKD and TF-QKD can be clearly revealed in the wave-state representation. In TF-QKD, Alice and Bob send wave-states, i.e. weak coherent states, to the third party Charlie for Bell state measurements\cite{lucamarini2018overcoming,yin2019measurement}. While in SEPM-QKD, Charlie sends the wave-entangled states to Alice and Bob to construct non-localized quantum correlations. In time order, the quantum state transmission and measurement of the two protocols are completely opposite. The key generation in both protocols comes from the wave-state correlation between Alice and Bob. According to the above analysis, the single-photon entanglement-based PM-QKD protocol is described as follows.

\par $\mathit{State}$ $\mathit{preparation}$. A single-photon state from a third untrusted party Charlie is sent to a 50:50 optical beam splitter to produce a single-photon entangled state close to the maximum entanglement. Next, he sends the photon states to Alice and Bob through two identical fibers with the same transmittance $\eta$. Because of channel noise and Eve's possible attack, the photon states reaching the terminals of Alice and Bob are not restricted to ideal single-photon entanglement. 

\par $\mathit{Selection}$ $\mathit{of}$ $\mathit{measurement}$ $\mathit{settings}$. With different phase-locking methods\cite{ma2012alternative,santarelli1994heterodyne}, the lase source of Alice and Bob are perfectly locked to achieve athe same global phase. Alice generates a random bit string $K_{a}$ in which each bit takes value $k_{a} \in \left \{ 0,1 \right \}$ and a random phase $\phi_{a} \in \left \{ -\frac{\pi}{4},0,\frac{\pi}{4},\frac{\pi}{2} \right \}$ corresponding to the measurements $\left( \sigma _{Z}-\sigma _{Y} \right )/\sqrt{2}$,  $\sigma _{Z}$, $\left (\sigma _{Z}+\sigma _{Y}\right )/\sqrt{2}$, $\sigma _{Y}$ and then prepares the corresponding weak coherent state $\left | \gamma e^{i\left ( \phi_{a} +k_{a}\pi \right )} \right \rangle$. Simultaneously, Bob generates a weak coherent state $\left | \gamma e^{i\left ( \phi_{b} +k_{b}\pi \right )} \right \rangle$ in which $k_{b} \in \left \{ 0,1 \right \}$ and $\phi_{b} \in \left \{ -\frac{\pi}{4},0,\frac{\pi}{4},\frac{\pi}{2} \right \}$. Alice and Bob interfere their weak coherent states with the single-photon state distributed by Charlie to measure the wave states and the interference results are recorded as the joint counting of the single-photon detectors on both sides.

\par $\mathit{Announcement}$. When all measurements are completed, Alice and Bob announce their detection results, i.e., the ordinal numbers of the fired single-photon detectors, and the phase values $\phi_{a}$ and $\phi_{b}$.

\par $\mathit{Sifting}$. A successful detection event is defined as having only one detector response on both sides at a given time. After they have announced the phases $\phi_{a}$ and $\phi_{b}$, the secret key is extracted when $\phi_{a}=\phi_{b}$. If the sum of the ordinal number $i+j$ is an even number, Alice and Bob keep their raw key; if $i+j$ is an odd number, then Bob flips his key.

\par $\mathit{Parameter}$ $\mathit{estimation}$. With a single-photon entanglement distribution, a bit-flipping error on the $X$ basis can never happen, otherwise photon number conservation is violated. In addition to entanglement degradation caused by channel transmission loss, information loss is mainly caused by phase noise, i.e., bit flipping on $Z$ and $Y$ bases. During the measurement, the selection of the $Z$ and $Y$ bases is equivalent, so the bit error rates on the two bases, $e_{Z}$ and $e_{Y}$, are equal.  Alice and Bob agree on a random bit string with half the length of the sifted key to be check-bit to measure the bit error rate $e$. Next, they use part of the remaining data in which $\left | \phi_{a}-\phi_{b} \right |=\frac{\pi}{4}$ to construct the Bell function $S$ on the $ZY$ plane to estimate the maximal information that may have leaked to Eve.

\par $\mathit{Key}$ $\mathit{distillation}$. In the post-processing, Alice and Bob perform error corrections in accordance with the bit error rate $e$ and privacy amplification according to the Bell function $S$ to generate the final secret key.

\section{Security of SEPM-QKD}

\par In SEPM-QKD, the key is distributed through a non-localized single-photon entangled state. Alice and Bob measure entangled states jointly. When entangled states are eigenstates of joint measurement operators, their measurements are perfectly correlated. They can extract keys based on joint measurement results or measurement settings. Eve's attack can be monitored based on violations of Bell's inequality. At first glance, this protocol belongs to DI-QKD. Although conventional DI-QKD is secured in theory, it is nevertheless difficult to distribute keys over long distances due to detection loopholes. 

\par Here, we point out that the detection loophole in the standard Bell experiment will not be a factor affecting the key security of the protocol. Previously, it was found that the security of QKD can be related to entanglement purification\cite{lo1999unconditional,shor2000simple}. The amount of security information that can be extracted between Alice and Bob is determined by the amount of purifiable entanglement. In DI-QKD, we certify that the bound of the accessible private key is determined by how much entanglement we can distill from the imperfect entangled state\cite{acin2007device,masanes2011secure,lim2013device}. 

\par In a standard Bell experiment, to give a rigorous proof of quantum delocalization, all loopholes in the experiment need to be closed, including the efficiency of the detector and transmission loss\cite{shalm2015strong}. For the DI-QKD protocol, we just need to accept quantum delocalization as rigorous and correct. After solving this issue, DI-QKD is equivalent to the BB84 protocol. In this protocol, we only focus on the data that can be measured successfully. In a conventional Bell experiment with polarization entanglement, the measurement in the $Z-$ and $X-$ bases needs the switching of the angle of the polarizers, which must be perfectly correlated with the secret key. This may leave Eve a chance to fabricate the measurement settings if she takes full control of the measurement setup. In the following, we need to establish whether in such an event Eve could fabricate a fake result of the Bell's inequality test given the limited information publicly announced by Alice and Bob.

\par In SEPM-QKD, Alice and Bob encode the key information in the phases of the weak coherent states. The encoding is equivalent to the measurement settings, and no switch of the measurement basis is needed. If this initial key information had been leaked to Eve, all QKD protocols would fail. From Eq. (5), the quantum measurement of the protocol may be considered to consist of three systems: the single-photon entangled state $\left | \Phi_{ A,B}^{-} \right \rangle$, the joint states of the single-photon detector $D$, and the corresponding joint key states $K$. The initial state of the total system is written
\begin{equation}
\rho_{\left( A,B \right)DK}=\rho_{A,B} \left |  N_{in} \right \rangle \left \langle N_{in} \right | \left |  \kappa_{in} \right \rangle \left \langle \kappa_{in} \right | ,
\end{equation}
 which is a tensor product of the three subsystems, with $\rho_{A,B}$ the single-photon entangled state sent by Charlie, and $\left | N_{in} \right \rangle$ and $\left |  \kappa_{in} \right \rangle$ the initial joint states of the two-sided single-photon detectors and the key state with $N=i+j$ and $\kappa=\left | k_{a}-k_{b} \right |$. Measurement is in general regarded as a unitary operation of the system; the joint measurement performed by Alice and Bob with two POVM elements $\left \{ E_{\kappa} \right \}$ may be written as
\begin{equation}
\varepsilon \left ( \rho_{\left( A,B \right)DK} \right )=  E_{0}^{+} \rho_{A,B} E_{0} \left |  even \right \rangle \left \langle even \right |  \left | 0 \right \rangle \left \langle 0 \right | 
+E_{1}^{+}\rho_{A,B} E_{1} \left |  odd \right \rangle \left \langle odd \right |  \left | 1 \right \rangle \left \langle 1 \right |.
\end{equation}
Once the QKD-protocol is determined, after the announcement of $N$ publicly, the information of $\kappa$ may be revealed by Eve. However, she still does not know the exact value of $k_{a}$ and $k_{b}$. At this stage, we find SEPM-QKD is equivalent to MDI-QKD. Eve barely gets any information about the measurement settings of Alice and Bob, so it is almost impossible for her to successfully fabricate the measurement results to cheat Alice and Bob.

\par With the presence of channel transmission losses and the imperfections in detection, Eve has the opportunity to implement various attack schemes. Even though a purification scheme for single-photon entanglement regarding phase noise have been provided\cite{sangouard2008purification,salart2010purification}, the reality is more complicated. Alice and Bob's extractable fully secure key rate has a lower bound given by\cite{acin2007device,devetak2005distillation,cai2009finite}
\begin{equation}
r\geq I\left ( A:B  \right )-\chi \left ( AB:E  \right ) ,
\end{equation}
where $I\left ( A:B  \right )=H\left ( A \right )-H\left ( A|B  \right )$ is the mutual information between Alice and Bob, which is equal to $1-H\left ( e \right )$, and $\chi \left ( AB:E \right )=S\left ( \rho_{AB|i,j} \right )-\sum_{c}p\left ( c \right ) S\left ( \rho_{AB|i,j}^{c} \right )$ the Holevo quantity between Eve and Alice and Bob after the ordinal numbers $i,j$ have been announced publicly, here, the quantity $H\left ( e \right )$ is the amount of information loss due to bit flipping errors, and $\chi \left ( AB:E \right )$ is the maximum amount of information Eve obtains from $\rho_{AB}$ at a given error rate $e$, and for values of $i,j$ and $\phi_{a},\phi_{b}$. 

\par There are two kinds of attack schemes on Alice and Bob that Eve could implement; they correspond to the two Holevo quantities $\chi \left ( AB:E  \right )$. One is a collective attack in which Eve correlates her system with the joint system of Alice and Bob and produces a total quantum state $\rho_{ABE}$. In this protocol, Eve can not get any information about the measurement settings, so she can't control the measurement process effectively. Her only freedom is to generate the joint quantum state, in which the results of Alice and Bob's reduced states are consistent with predictions from theory, taking into account the imperfections in the equipment.

\par  Under the idea of coherent attack, Eve uses weak measurements to obtain information of quantum states. The limitation of the attack is that the delocalized quantum correlation between Alice and Bob is within the acceptable range of them. For uniformly random marginals in the $ZY$ plane, Eve's maximal collective attack will be saturated by sending the entangled single-photon state of which he holds a purification\cite{acin2007device}
\begin{equation}
\left | \Psi_{ABE} \right \rangle=\frac{1}{2}\left( I+H_{A}H_{B} \right)  [ \sqrt{1-2e} \left | E_{0} \right \rangle \left | \Phi_{AB}^{-} \right \rangle_{z}
 +\sqrt{2e}\left | E_{1} \right \rangle \left | \Phi_{AB}^{+} \right \rangle_{z} ],
\end{equation}
where $I$ is the identity density operator, $H_{A}$ and $H_{B}$ are Hadamard matrices operated on Alice's and Bob's wave states in $ZY$ plane, which transform $Z$ basis to $Y$ basis, $\left | E_{0}\right \rangle$ and $\left | E_{1}\right \rangle$ are the two orthogonal states hold by Eve, $\left | \Phi_{AB}^{\pm} \right \rangle_{z}$ are the Bell states under the representation of $Z$ basis. A simple derivation of Eve's maximum collective attack is given in the method section. We find that the maximum violation of the CHSH-Bell inequality is $S=2\sqrt{2}\left( 1-2e \right)$. 

\par We readily find that $\chi_{1} \left( AB:E \right)\leq 2e$, which means that whenever Alice and Bob negotiate one bit of information, Eve can successfully steal $2e$ bit of information. Next, we examine the scope of the Bell-inequality verification. Assume that Eve intercepts the single-photon entangled state and induce a certain amount of error rate. The maximum error rate that Bell inequality tolerates is $14.6\%$, which is larger than $11\%$\cite{shor2000simple}, the maximum error rate that Alice and Bob can tolerate in extracting finite information against Eve's collective attacks. Therefore, violation tests of Bell's inequality violation are a feasible scheme for monitoring Eve's collective attack.

\par The other possible attack scenario for Eve is the beam-splitting (BS) attack, in which the loss of a single-photon entangled state in optical channels can be considered to be stored by Eve and measured after Alice and Bo have announced publicly their measurement basis and random phase, as shown in Fig. 4. Thus, the BS attack is an individual attack that is independent of a collective attack and can not be found with Bell's inequality tests. Considering channel loss, the single-photon state between Alice, Bob, and Eve is written
\begin{equation}
\left | \Psi_{ABE} \right \rangle=\frac{\sqrt{2}}{2} [\sqrt{\eta} \left ( \left | 1_{A}0_{B} \right \rangle + \left | 0_{A}1_{B} \right \rangle \right )\left | 0_{E_{A}}0_{E_{B}} \right \rangle 
+ \sqrt{1- \eta } \left | 0_{A}0_{B} \right \rangle \left ( \left | 1_{E_{A}}0_{E_{B}} \right \rangle + \left | 0_{E_{A}}1_{E_{B}} \right \rangle \right ) ],
\end{equation}
which is a single-photon multi-mode asymmetric W-state\cite{heaney2011extreme,sheng2014protecting}, where $\frac{\sqrt{2}}{2} \left ( \left | 1_{E_{A}}0_{E_{B}} \right \rangle + \left | 0_{E_{A}}1_{E_{B}} \right \rangle \right )$ is the state responsible for channel loss, which is assumed to be stored by Eve, whose system is entangled with the systems of Alice and Bob. Suppose Eve uses weak coherent light of the same intensity as Alice and Bob to measure the wave state. After Alice and Bob announce their random phases $\phi_{a}$, $\phi_{b}$ as well as the ordinal numbers $i,j$ of the single-photon detectors, for a given channel transmittance $\eta$ and local coherent field amplitude $\gamma$, the maximum information that Eve can gain from Alice and Bob is
\begin{equation}
\chi_{2} \left ( AB:E \right )=\frac{\gamma ^{4}\left ( 1+3\eta +2\gamma^{2}  \right )}{4}\left[1-H\left( p \left( \eta,\gamma \right) \right )\right],
\end{equation}
where the quantity $p \left( \eta, \gamma \right)$ is the normalized probability that Eve uses to guess the key of Alice and Bob; its expression is
\begin{equation}
p \left( \eta \right)=  \frac{1+3\eta -4\sqrt{\eta \left ( 1-\eta  \right )}+2\gamma^{2}}{2+6\eta+4\gamma^{2} }.
\end{equation}
See the derivation in the method section. Now, if the BS attack is not considered, the key rate in Eq. (8) is found to be equal to the amount of entanglement that can be distilled between Alice and Bob. This security proof is equivalent to the security proof of BB84 QKD based on entanglement purification\cite{lo1999unconditional,shor2000simple}. The loss of these two parts of the information corresponds to an error correction and private amplification in post-processing. After considering Eve's two attack schemes, the lost information for private amplification should be recalibrated.

\section{Simulation and discussion}

\par Next, we simulate the distance-dependent key rate in a practical situation. Among all the successful detection events, there are three kinds of false detection events, which constitute the detection error rate $e$. These events come from dark counting of detectors, phase insensitive interference, and phase misalignment. For all single-photon detectors with the same dark count rate $p_{dark}$, the rate of successful detection events $p_{r,dark}$ and false detection events $p_{e,dark}$ caused by dark counting are both equal to $2p_{dark}^{2}$. For the joint measurement of wave states, there is a small portion of detection events stemming from phase-insensitive interference, a HOM-type of interference. The rate for joint HOM interference is $p_{HOM}=\gamma^{4}\eta_{d}^{2}/4$ with $\eta_{d}$ the detection efficiency of the single-photon detectors, and gives rise to a correct detection rate $p_{r,HOM}=\gamma^{4}\eta_{d}^{2}/8$ and a false detection rate $p_{e,HOM}=\gamma^{4}\eta_{d}^{2}/8$. In the last false detection event, the misalignment error rate is $e_{d}$, the contribution to the total error rate being $p_{d}e_{d}$, where $p_{d}=\gamma^{2}\eta_{d}^{2}\eta/2$ is the probability of a joint measurement of wave states in ideal single-photon entanglement. Then the error rate $e$ in terms of these parameters is expressed as
\begin{equation}
e\approx  \frac{p_{e,dark}+p_{e,HOM}+p_{d}e_{d}}{p_{dark}+p_{HOM}+p_{d}}.
\end{equation}

\par After taking into account all practical factors, such as error correction and privacy amplification, we obtain a final lower bound of the key rate of
\begin{equation}
r\geq Q\left [1-fH\left ( e \right )-2e-\frac{\gamma^{2}\left ( 1+3\eta+2\gamma^{2} \right )}{\eta}\left[1-H\left( p \left( \eta,\gamma \right) \right )\right]\right ],
\end{equation}
where $Q=p_{dark}+p_{HOM}+p_{d}$ is the rate of the joint measurement of the wave states,  $\eta=\exp\left( -\alpha_{f} x \right )$ the channel transmittance with $\alpha_{f}$ the coefficient of absorption and $x$ the transmission distance, and $f$ the inefficiency of error correction, which always takes the value between $1.2$ and $2$ in accordance with the error correction protocol\cite{lo2012measurement}. In this formula, we have assumed the transmittance of the optical fibers, the amplitude of the local oscillator fields, and the detector efficiency are the same for Alice and Bob.

\par The simulation results of our SEPM-QKD under different intensities of local coherent fields is shown in Fig.2. The coefficient of transmission loss for the optical fiber at 1550 nm is $\beta_{l}=0.2dB/km$ and the coefficient of absorption is $\alpha_{f}=\left( \beta_{l}\ln10 \right)/10$. Also, the detection efficiency at this frequency $\eta_{d}$ is $14.5\%$ for a commercial single-photon detector, the dark count rate is $p_{dark}=8\times 10^{-8}$ for all detectors, and the misalignment error $e_{d}$ is $1.5\%$\cite{ma2018phase}, the value for the inefficiency of error correction is set at $f=1.2$\cite{lo2012measurement}. From this figure, the key rate is seen to that the key rate decrease as the intensity of the local coherent light field decreases; because the probability of successful joint detection events is lower as the amplitude $\gamma$ decreases. However, the transmission distance shows an opposite trend in its dependence on intensity. The dependence of the transmission distance on the amplitude $\gamma$ arises from the false detection of phase insensitive joint counts $p_{HOM}$, which is proportional to the square of the light intensity, yielding $\gamma^{4}$. For a specified QKD protocol, the transmission distance is a compromise between the signal rate and the error rate. As the amplitude $\gamma$ decreases, the phase-insensitive joint count-induced error plays little role in the key distillation. Therefore, a longer transmission distance obtains. We also compare the performance of SEPM-QKD for $\gamma=0.002$ and $\gamma=0.001$ with and without BS attacks. For short transmission distance, BS attack has negligible effect on the key rate, but for long transmission distance, the effect of BS attack should not be ignored.

\par Here, we make a clear comparison between different QKD protocols (Fig.3), in which $\gamma=0.001$ is chosen for the SEPM-QKD scheme. For traditional single-photon based BB84- and MDI-QKD schemes, their key rate obey the well-known linear bound by Pirandola et al. (PLOB bound)\cite{pirandola2017fundamental}. However, we see that, like PM-QKD and TF-QKD, SEPM-QKD displays a quadratic increase in the key rate with respect to the transmission distance which obey the single-repeater bound\cite{pirandola2019end}. For short transmission distances, the key rate of SEPM-QKD is not only less than that of PM-QKD and TF-QKD, but also lower than particle-state based QKDs, like BB84- and MDI-QKD. There are two reasons for this result. The first reason is that the average intensity of the light source in SEPM-QKD is far lower than all other QKD protocols. The second reason is that BS attack is considered in SEPM-QKD, but not in other protocols. 

\par It can clearly be seen that SEPM-QKD clarifies in principle the essential difference between TF-QKD and its variants which violate PLOB linear bound and BB84- and MDI-QKD proposed previously. In these QKD protocols, due to the different properties of information carrier and the different quantum states for distributing quantum keys, their implementation also has different technical challenges. In SEPM-QKD, a single-photon source is needed to generate wave-state entanglement. The single-photon produced in current experiments is probabilistic, which will reduce the quantum correlation between Alice and Bob. Under the current technical conditions, the heralded single-photon source is an effective solution to this problem. In addition, we can see from Eq. (5) that the phase insensitive interference, i.e. particle space interference, exists in coincidence counting, which results in the inability to use strong light intensity in SEPM-QKD, and the key rate is much lower than other QKD protocols. In our future work, we will propose a de-localized detection scheme to the performance of SEPM-QKD.

\section{Conclusion}

\par We have reported a phase matching QKD based on single-photon entanglement. This SEPM-QKD is a time-reversed version of TF-QKD, in which the secret key is encoded in wave space characterized by the phase value. Measurement settings in SEPM-QKD, like quantum keys, are encoded in the phase of the locally coherent state, so the detection loophole is closed. This contrasts that for conventional DI-QKD. For a given light source intensity, just like TF-QKD, SEPM-QKD improves the bound of key rate from $O\left( \eta \right )$ to $O\left(  \sqrt{\eta}\right )$. In the proof of security, we find that BS attacks will have a significant impact on the performance of the protocol for long-distance transmission. By comparison with single-photon QKD schemes, we found that in SEPM-QKD and TF-QKD the wave state can be used as a new information carrier that has different properties due to interference-induced detection enhancement, which allows photons to travel in fibers without obeying the PLOB bound. In the future, we wish to reduce the impact of the phase-insensitive coincidence counting rate on the key rate and to improve the key rate and transmission distance of SEPM-QKD.

\section*{Methods}

We present the methods for deriving the key rate formula in the main text. These methods theoretically give the upper limit of key rate obtained by Eve under the eavesdropping scheme of collective attack and beam-splitting attack. 

\textbf{Collective attack.} Collective attack is considered to be the most powerful side-channel attack through using the imperfection of Alice and Bob's experimental devices. Eve's attack operation must obey the law of quantum mechanics, and the bit error rate between Alice and Bob caused by eavesdropping should be within the predetermined range. Under the idea of collective attack, Eve obtains the quantum state information shared between Alice and Bob as much as possible through weak measurements. In BB84 protocol, collective attack can be described as Eve attaching his probe to each of the states sent by Alice to Bob, and performing unitary operation, so that his probe can be quantum correlated with the transmitted quantum states\cite{biham1997security,biham2002security}. Suppose that the interaction occurs in a two-dimensional space formed by a pair of orthogonal states $\left | p \right \rangle$ and $\left | q \right \rangle$. Eve's initial quantum state is $\left | E \right \rangle$, the interaction is represented by unitary operator $U$, 
\begin{equation}
U \left | E \right \rangle \left | p \right \rangle = \left | E_{p} \right \rangle \left | \alpha_{p} \right \rangle, U \left | E \right \rangle \left | q \right \rangle = \left | E_{q} \right \rangle \left | \beta_{q} \right \rangle,
\end{equation}
where $\alpha$ and $\beta$ are the rotation angles of the transmitted quantum states with respect to $\left | p \right \rangle$ and $\left | q \right \rangle$, respectively, $\left | E_{p} \right \rangle$ and $\left | E_{q} \right \rangle$ are the corresponding states owned by Eve. According to the unitarity of operator , we have the following equality
\begin{equation}
\left \langle E | E \right \rangle \left \langle p | q \right \rangle =0= \left \langle p | \left \langle E | U^{+} U | E \right \rangle | q \right \rangle = \left \langle E_{p} | E_{q} \right \rangle \left \langle \alpha_{p} | \beta_{q} \right \rangle.
\end{equation}
For any quantum states $\left | E_{p} \right \rangle$, $\left | E_{q} \right \rangle$, we have $\left \langle \alpha_{p} | \beta_{q} \right \rangle=0$. Therefore, the quantum states $\left | p \right \rangle$ and $\left | q \right \rangle$ are rotated at the same angle under weak measurements. Eve can reverse-rotate the transmitted quantum state after the unitary operation, 
\begin{equation}
T \left | E \right \rangle \left | p \right \rangle =RU \left | E \right \rangle \left | p \right \rangle = \left | E_{p} \right \rangle \left | p \right \rangle,  T \left | E \right \rangle \left | q \right \rangle=RU \left | E \right \rangle \left | q \right \rangle = \left | E_{q} \right \rangle \left | q \right \rangle.
\end{equation}
The relation between $\left | p \right \rangle$, $\left | q \right \rangle$ and the bases in $Z$ space and $X$ space can be written as
\begin{equation}
\begin{split}
\left | p \right \rangle &= a \left | 0 \right \rangle +b \left | 1 \right \rangle, \left | q \right \rangle= b \left | 0 \right \rangle -a \left | 1 \right \rangle;\\
\left | p \right \rangle&= \frac{\sqrt{2}}{2}\left [\left ( a+b \right ) \left | + \right \rangle +\left ( a-b \right ) \left | - \right \rangle  \right ],  \left | q \right \rangle= \frac{\sqrt{2}}{2}\left [\left ( a+b \right ) \left | - \right \rangle - \left ( a-b \right ) \left | + \right \rangle  \right ],
\end{split}
\end{equation}
where coefficients $a$ and $b$ satisfy normalization conditions $a^{2}+b^{2}=1$.
Then the weak measurements on these states can be described as
\begin{equation}
\left\{\begin{matrix}
T\left | E \right \rangle \left | 0 \right \rangle =\left | 0 \right \rangle \left ( a^{2}\left | E_{p} \right \rangle+b^{2}\left | E_{q} \right \rangle \right )+\left | 1 \right \rangle \left ( ab\left | E_{p} \right \rangle-ab\left | E_{q} \right \rangle \right ),\\ 
T\left | E \right \rangle \left | 1 \right \rangle =\left | 0 \right \rangle \left ( ab\left | E_{p} \right \rangle-ab\left | E_{q} \right \rangle \right )+\left | 1 \right \rangle \left ( b^{2}\left | E_{p} \right \rangle+a^{2}\left | E_{q} \right \rangle \right );
\end{matrix}\right.
\end{equation}
and
\begin{equation}
\left\{\begin{matrix}
T\left | E \right \rangle \left | + \right \rangle =\frac{1}{2}\left [\left | + \right \rangle \left ( \left (a+b \right )^{2}\left | E_{p} \right \rangle+\left (a-b \right )^{2}\left | E_{q} \right \rangle \right )+\left | - \right \rangle \left ( a^{2}-b^{2} \right )\left ( \left | E_{q} \right \rangle-\left | E_{p} \right \rangle \right )  \right ],\\ 
T\left | E \right \rangle \left | - \right \rangle =\frac{1}{2}\left [\left | + \right \rangle \left ( a^{2}-b^{2} \right )\left ( \left | E_{q} \right \rangle-\left | E_{p} \right \rangle \right )+\left | - \right \rangle \left ( \left (a-b \right )^{2}\left | E_{p} \right \rangle+\left (a+b \right )^{2}\left | E_{q} \right \rangle \right )  \right ].
\end{matrix}\right.
\end{equation}
After all the unitary operations, Eve perform unambiguous discrimination measurements on his states $\left | E_{p}  \right \rangle$ and $\left | E_{q}  \right \rangle$ to obtain the information between Alice and Bob. Alice and Bob's pre-agreed system bit error rate is $p_{e}$. A bit error occurs when Alice sends state $\left | 0 \right \rangle $ and Bob receives $\left | 1 \right \rangle $ or Alice sends state $\left | 1 \right \rangle $ and Bob receives $\left | 0 \right \rangle $. The total bit error rate is bounded by
\begin{equation}
1-\left \langle E_{p}|E_{q} \right \rangle=4p_{e}.
\end{equation}
We will demonstrate in our forthcoming paper that Eve could obtain the largest information when he sets $a=0$ and $b=1$ or $a=1$ and $b=0$. Eve's maximal information is bounded by $\frac{1-\left \langle E_{p}|E_{q} \right \rangle}{2}=2p_{e}$. 

According to conclusion from BB84, Eve could also performs the same collective attack on entangled state. Suppose Alice and Bob share singlet state $\Phi_{AB}^{-}$. Eve's collective attack on the entangled state can be formulated as
\begin{equation}
T\left | \Phi_{AB}^{-} \right \rangle \left | E_{A} \right \rangle \left | E_{B} \right \rangle=\frac{\sqrt{2}}{2}\left [ \left | 0 \right \rangle_{A}\left | 0 \right \rangle_{B} \left | E_{0} \right \rangle_{A}\left | E_{0} \right \rangle_{B}-\left | 1 \right \rangle_{A}\left | 1 \right \rangle_{B} \left | E_{1} \right \rangle_{A}\left | E_{1} \right \rangle_{B}\right ].
\end{equation}
Here we may set the joint state $\left | E_{0} \right \rangle_{A}\left | E_{0} \right \rangle_{B}$ to $\left | E_{0} \right \rangle_{Z}$, and the joint state $\left | E_{1} \right \rangle_{A}\left | E_{1} \right \rangle_{B}$ to $\left | E_{1} \right \rangle_{Z}$. A pair of orthogonal bases can be constructed from these two states
\begin{equation}
\begin{split}
\left | E_{\parallel } \right \rangle &=\frac{1}{\sqrt{2\left ( 1+\left \langle E_{0}|E_{1} \right \rangle_{Z} \right )}}\left [ \left | E_{0 } \right \rangle_{Z}+ \left | E_{1 } \right \rangle_{Z} \right ],\\
\left | E_{\perp  } \right \rangle &=\frac{1}{\sqrt{2\left ( 1-\left \langle E_{0}|E_{1} \right \rangle_{Z} \right )}}\left [ \left | E_{0 } \right \rangle_{Z}- \left | E_{1 } \right \rangle_{Z} \right ].
\end{split}
\end{equation}
Thus, under collective attack, the joint quantum state between Eve and Alice and Bob is
\begin{equation}
\begin{split}
\left | \Psi_{ABE } \right \rangle &=\frac{\sqrt{2}}{2}\left [ \left | 0 \right \rangle_{A}\left | 0 \right \rangle_{B} \left | E_{0} \right \rangle_{Z}-\left | 1 \right \rangle_{A}\left | 1 \right \rangle_{B} \left | E_{1} \right \rangle_{Z}\right ]\\
&=\frac{\sqrt{2}}{2}\left [ \sqrt{1+\left \langle E_{0}|E_{1} \right \rangle_{Z}}\left | \Phi_{AB}^{-} \right \rangle\left | E_{\parallel } \right \rangle +\sqrt{1-\left \langle E_{0}|E_{1} \right \rangle_{Z}}\left | \Phi_{AB}^{+} \right \rangle \left | E_{\perp  } \right \rangle \right ].
\end{split}
\end{equation}
From Eq. (23), we can find that Eve could steals $1-\left \langle E_{0}|E_{1} \right \rangle_{Z}$ information in $Z$ space without causing any bit flipping error, while he steals nothing in $X$ space bu causing a bit error rate of $\frac{1}{2}\left ( 1-\left \langle E_{0}|E_{1} \right \rangle_{Z} \right )$. In order to balance the bit error rate of $Z$ space and $Y$ space, Eve will rotate the transmitted entangled state, and get the result of Eq. (9) in the main text.

\textbf{Beam-splitting attack.}
In the beam separation attack, we assume that the transmission loss of a single photon is all intercepted and stored by Eve, and finally an asymmetric W-state between Eve and Alice and Bob is formed, as shown in Eq. (10). The details of BS attack is shown in Fig. 4. Compared with collective attack, beam-splitting attack can be regarded as a passive attack scheme. After Alice and Bob announce their bases publicly, Eve conducts a homodyne on the stored quantum state by using the coherent state with the same intensity as Alice and Bob. The joint detection can be expressed as
\begin{equation}
\begin{split}
\left | \Psi_{ABE} \right \rangle&=\left [ \left | 0 \right \rangle_{A}+\gamma e^{i\theta _{A}}\left | 1 \right \rangle_{A} \right ]\left [ \left | 0 \right \rangle_{B}+\gamma e^{i\theta _{B}}\left | 1 \right \rangle_{B} \right ]\left [ \left | 0 \right \rangle_{E_{A}}+\gamma e^{i\theta _{E_{A}}}\left | 1 \right \rangle_{E_{A}} \right ]\left [ \left | 0 \right \rangle_{E_{B}}+\gamma e^{i\theta _{E_{B}}}\left | 1 \right \rangle_{E_{B}} \right ]\\
&\times \frac{\sqrt{2}}{2}\left [ \sqrt{\eta }\left ( \left | 1_{A}0_{B} \right \rangle+\left | 0_{A}1_{B} \right \rangle \right )\left | 0_{E_{A}}0_{E_{B}} \right \rangle+\sqrt{1-\eta }\left | 0_{A}0_{B} \right \rangle\left ( \left | 1_{E_{A}}0_{E_{B}} \right \rangle+\left | 0_{E_{A}}1_{E_{B}} \right \rangle \right ) \right ],
\end{split}
\end{equation}
where $\left | 1_{E_{A}} \right \rangle$ and $\left | 1_{E_{B}} \right \rangle$ represent the photon states stored by Eve on Alice's and Bob's sides. 
The single-photon W-state then interferes with the coherent state on the beam splitter
\begin{equation}
\begin{split}
\left | \Psi_{ABE} \right \rangle&=\left [ \left | 0 \right \rangle+\gamma e^{i\theta _{A}}\frac{\sqrt{2}}{2}\left [ i\left | 1 \right \rangle_{A_{1}}+\left | 1 \right \rangle_{A_{2}} \right ] \right ]\left [ \left | 0 \right \rangle+\gamma e^{i\theta _{B}}\frac{\sqrt{2}}{2}\left [ i\left | 1 \right \rangle_{B_{1}}+\left | 1 \right \rangle_{B_{2}} \right ] \right ]\\
&\times \left [ \left | 0 \right \rangle+\gamma e^{i\theta _{E_{A}}}\frac{\sqrt{2}}{2}\left [ i\left | 1 \right \rangle_{E_{A1}}+\left | 1 \right \rangle_{E_{A2}} \right ] \right ]\left [ \left | 0 \right \rangle+\gamma e^{i\theta _{E_{B}}}\frac{\sqrt{2}}{2}\left [ i\left | 1 \right \rangle_{E_{B1}}+\left | 1 \right \rangle_{E_{B2}} \right ] \right ]\\
&\times \frac{1}{2} \left [ \sqrt{\eta }\left ( \left [ \left | 1 \right \rangle_{A_{1}}+i\left | 1 \right \rangle_{A_{2}} \right ]+\left [ \left | 1 \right \rangle_{B_{1}}+i\left | 1 \right \rangle_{B_{2}} \right ] \right )+\sqrt{1-\eta }\left ( \left [ \left | 1 \right \rangle_{E_{A1}}+i\left | 1 \right \rangle_{E_{A2}} \right ]
+\left [ \left | 1 \right \rangle_{E_{B1}}+i\left | 1 \right \rangle_{E_{B2}} \right ] \right ) \right ].
\end{split}
\end{equation}
When Alice, Bob and Eve have coincidence counts, Eve has a certain probability of stealing Alice and Bob's key information. Without losing generality, we assume that Alice and Bob's measurement bases are in Z space at one moment, and their fired detectors are $A_{1}$ and $B_{1}$, which means $\theta_{A}= \theta_{B}=\theta\in \left \{ 0, \pi \right \}$. By expanding Eq. (25), the terms satisfying the above conditions are
\begin{equation}
\begin{split}
&-\frac{\gamma^{2}}{4}\left [ \sqrt{1-\eta} e^{i\left ( \theta_{A}+ \theta_{B}\right )} + \sqrt{\eta}e^{i \theta_{E_{A}} }\left ( e^{i \theta_{A} }+e^{i\theta_{B}} \right )  \right ] \left | 1_{A_{1}}1_{B_{1}}1_{E_{A1}} \right \rangle;\\
&i\frac{\gamma^{2}}{4}\left [ -\sqrt{1-\eta} e^{i\left ( \theta_{A}+ \theta_{B}\right )} + \sqrt{\eta}e^{i \theta_{E_{A}} }\left ( e^{i \theta_{A} }+e^{i\theta_{B}} \right )  \right ] \left | 1_{A_{1}}1_{B_{1}}1_{E_{A2}} \right \rangle.
\end{split}
\end{equation}
When Eve synchronizes the light source with Alice and Bob's measurements, he randomly sets the value of $\theta_{E_{A}}$ to $0$ or $\pi$, and infers the value of $\theta$ from the detection results. Here by setting $\theta_{E_{A}}=0$, then we have the joint detection probabilities
\begin{equation}
\begin{split}
P\left ( \theta, E_{A1} \right )&=\frac{\gamma^{4}}{16}\left( 1+3\eta+4\sqrt{\eta \left ( 1-\eta \right )} \cos \theta  \right );\\
P\left ( \theta,E_{A2} \right )&=\frac{\gamma^{4}}{16}\left( 1+3\eta-4\sqrt{\eta \left ( 1-\eta \right )} \cos \theta \right ).
\end{split}
\end{equation}
There are phase-independent joint detection events between Alice, Bob and Eve, whose probability is equal to $\frac{\gamma^{6}}{8}$. The joint probability matrix between Alice and Bob's phase $\theta$ and Eve's detection results $E_{A}$ is
\begin{equation}
p\left ( \theta ,E_{A} \right )=\frac{\gamma ^{4}\left ( 1+3\eta +2\gamma^{2}  \right )}{8}\begin{bmatrix}
\frac{1+3\eta +4\sqrt{\eta \left ( 1-\eta  \right )}+2\gamma^{2}}{2+6\eta+4\gamma^{2} } & \frac{1+3\eta -4\sqrt{\eta \left ( 1-\eta  \right )}+2\gamma^{2}}{2+6\eta+4\gamma^{2}  } \\ 
 \frac{1+3\eta -4\sqrt{\eta \left ( 1-\eta  \right )}+2\gamma^{2}}{2+6\eta+4\gamma^{2} }& \frac{1+3\eta +4\sqrt{\eta \left ( 1-\eta  \right )}+2\gamma^{2}}{2+6\eta+4\gamma^{2}  },
\end{bmatrix}
\end{equation}
where $1+3\eta+2\gamma^{2}$ is the normalization constant. Of course, Eve can also carry out similar beam-splitting attacks on $E_{B}$ and get the same results.

\acknowledgments
This work is supported by Young fund of Jiangsu Natural Science Foundation of China (SJ216025), National fund incubation project (NY217024), Scientific Research Foundation of Nanjing University of Posts and Telecommunications (NY215034), the National Natural Science Foundation of China (No. 61475075), the open subject of National Laboratory of Solid State Microstructures of Nanjing University (M31021).

\bibliography{reference}
\section*{Figures}
\begin{figure}[ht]
\centering
\includegraphics[width=120mm]{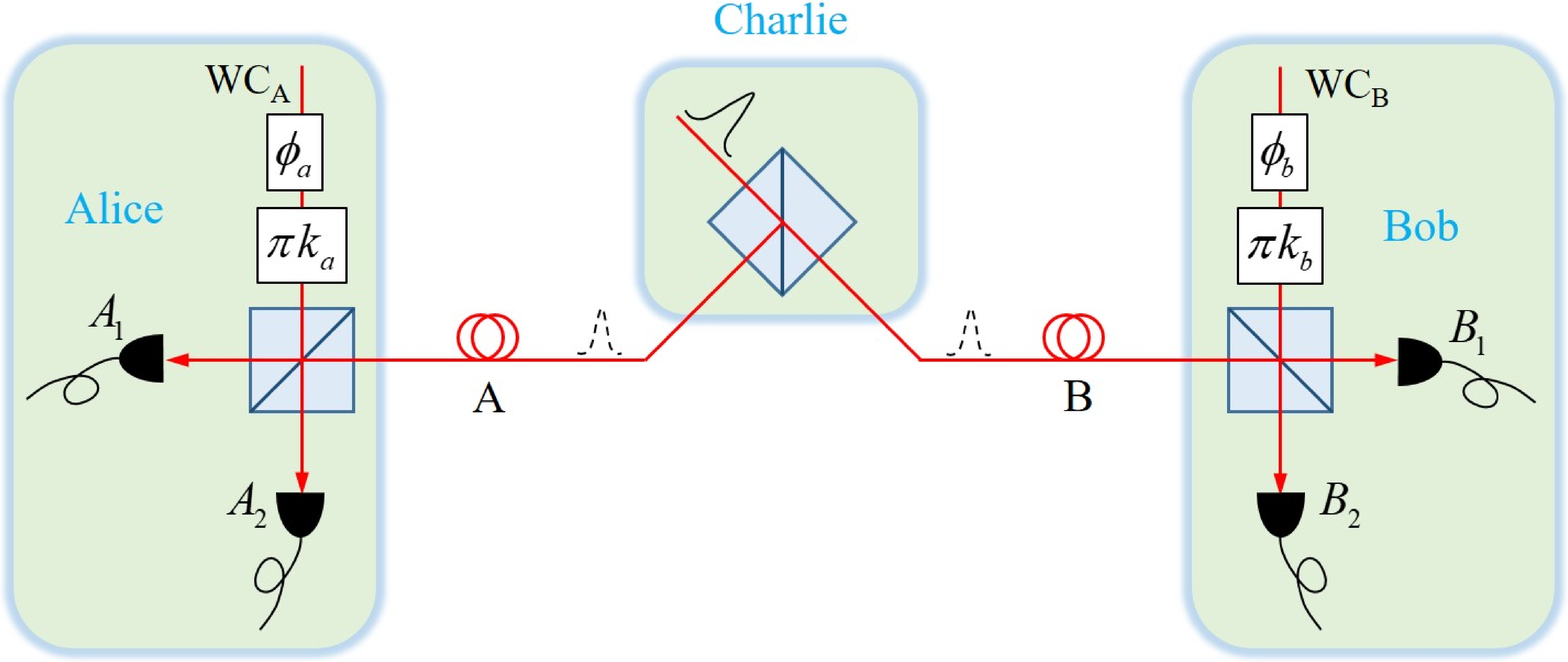}
\caption{Schematic diagram of SEPM-QKD. An untrusted third party, Charlie, generates single-photon entanglement, by injecting a photon from a heralded single-photon source into a beam splitter. Alice and Bob generate a local weak coherent (WC) state $\left | \gamma e^{i\left ( \phi_{a(b)}+k_{a(b)}\pi \right )} \right \rangle$ with $\phi_{a(b)}\in \left \{ -\frac{\pi}{4},0,\frac{\pi}{4},\frac{\pi}{2} \right \}$ and $k_{a(b)}\in \left \{ 0,1 \right \} $ to test the quantum nonlocal correlation in wave space and generate the final key. $\phi_{a(b)}$ is a random phase used to construct Bell inequality and is also used for phase matching measurement. Random bit $k_{a(b)}$ can be regarded as the measurement setup for homodyne detection of wave-state.}
\label{Fig. 1}
\end{figure}

\begin{figure}[ht]
\centering
\includegraphics[width=120mm]{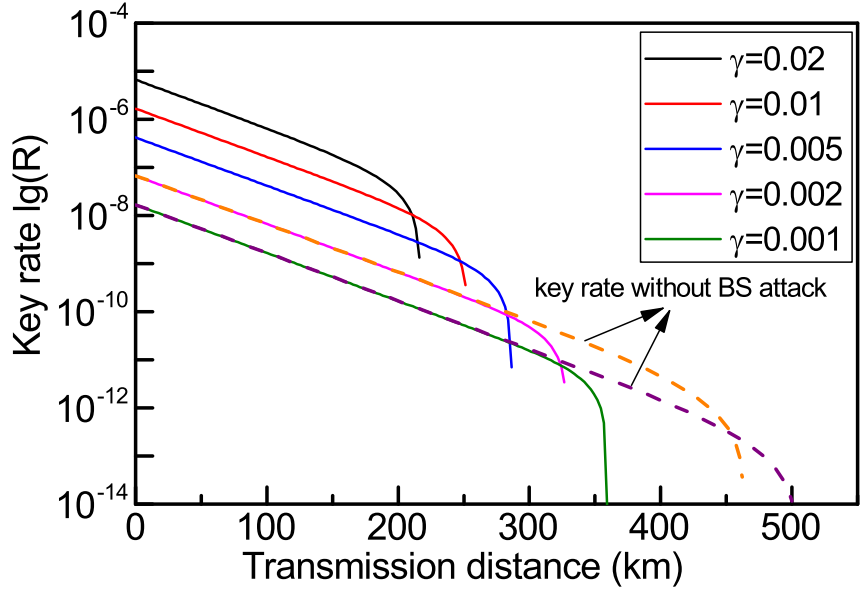}
\caption{Simulation of SEPM-QKD under intensities of local coherent light. The key rate decreases with increasing attenuation of the coherent light intensity whereas the transmission distance increases as the attenuation increases. When the average photon-number of the coherent state is far less than 1, the key rate is approximately proportional to the square of the amplitude of the coherent state according to Eqs. 14. For coherent states with high intensity, the proportion of the particle-like correlation between Alice and Bob will also increase (Eqs. 5). This will increase the bit error rate of the final key, so the transmission distance will be reduced. In addition, there are two more key rate curves, orange and purple dotted lines, which correspond to the fitting results without considering beam-splitting (BS) attacks. It can be found that the BS attack will have an important effect on the key rate at long transmission distance.}
\label{Fig. 2}
\end{figure}

\begin{figure}[ht]
\centering
\includegraphics[width=120mm]{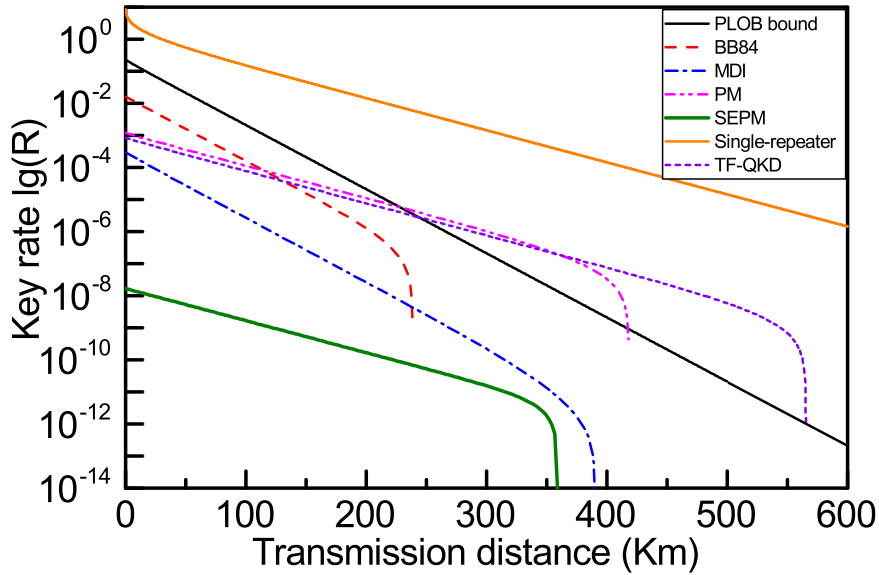}
\caption{Key rate comparison between different QKD protocols. The simulation results of the other QKD are taken from Ref.\cite{lucamarini2018overcoming,ma2018phase}. Compared with single-photon based BB84- and MDI-QKD schemes which obey the PLOB bound by Pirandola et al. (PLOB bound)\cite{pirandola2017fundamental}, SEPM-QKD has the same $\sqrt{\eta}$ dependence on transmission distance as PM-QKD and TF-QKD which obey the single-repeater bound\cite{pirandola2019end}. In BB84- and MDI-QKD protocols, the carrier of information is a single-photon, and the detection probability is proportional to the transmission coefficient $\eta$, so the key rate has a $\eta$ on the transmission distance. While for PM- and SEPM-QKD protocols, the carrier of information is a wave-photon, and the detection probability is proportional to the square root of the transmission coefficient $\sqrt{\eta}$, so the key rate has a $\sqrt{\eta}$ on the transmission distance. In the simulation of SEPM-QKD, the amplitude of coherent state is $\gamma=0.001$. Its average intensity is several orders of magnitude lower than that in other QKD protocols and BS attack is also considered in SEPM-QKD, which results in the key rate of SEPM-QKD being much lower than that of other QKD protocols.}
\label{Fig. 3}
\end{figure}

\begin{figure}[ht]
\centering
\includegraphics[width=120mm]{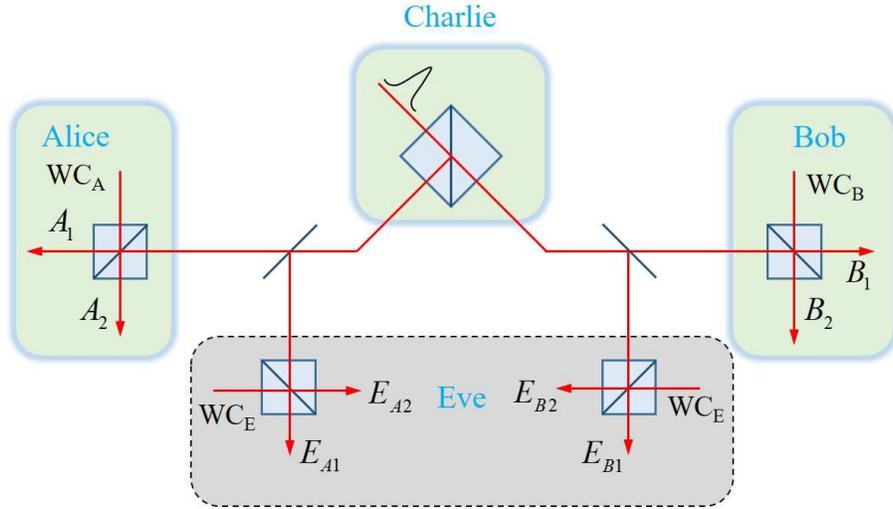}
\caption{Schematic diagram of BS attack. Suppose that the transmission loss of a single-photon state is captured and stored by Eve in BS attack scheme. In this attack scheme, Eve synchronizes his light source with Alice and Bob's. After Alice and Bob publicly announce random phase values, selection of measurement bases and response results of detectors, Eve uses the same measurement method to measure the stored photon states. Eve finally infers Alice and Bob's keys based on his measurement result $E_{A,B}$.}
\label{Fig. 1}
\end{figure}
\end{document}